# Achievable Degrees of Freedom of the K-user Interference Channel with Partial Cooperation

Ahmed A.Naguib, Khaled Elsayed
EECE Department, Cairo University
a.abdulkarim,khaled@ieee.org

Mohammed Nafie
WINC Lab, Nile University
mnafie@nileuniversity.edu.eg

*Abstract*—In this paper, we consider the $K$-user interference channel with partial cooperation, where a strict subset of the $K$ users cooperate. For the $K$-user interference channel with cooperating subsets of length $M$, the outer bound of the total degrees of freedom is $KM/(M+1)$. In this paper, we propose a signal space-based interference alignment scheme that proves the achievability of these degrees of freedom for the case $K = M + 2$. The proposed scheme consists of a design for the transmit precoding matrices and a processing algorithm which we call the Successive Interference Alignment (SIA) algorithm. The decoder of each message uses the SIA algorithm to process the signals received by the M cooperating receivers in order to get the maximum available degrees of freedom.

*Index Terms*—Degrees of freedom, interference alignment, cooperation.

## I. INTRODUCTION

Interference between users arises as one of the most challenging problems for multi-user wireless communication systems. The interference channel (IC) models the network in which a number of transmitter-receiver pairs communicating simultaneously, and each transmitter has one message to only one of the receivers. The simplest possible IC, the 2-user IC, is extensively studied over more than thirty years and several inner and outer bounds have been developed. However, the capacity of this channel in general is still an open problem and the best results known so far are derived using approximation [1]. Recently, the approach of degrees of freedom paved the way for developing schemes that make a revolution on how to manage interference in wireless networks. The degrees of freedom (DoF) approximate the capacity of the network as:

$$C(SNR) = d \log(SNR) + O(\log(SNR))$$

where $C(SNR)$ is the capacity of the network at signal to noise ration equals to $SNR$, $d$ is DoF available in the network. At high $SNR$, $O(log(SNR))$ term becomes negligible compared with $log(SNR)$. Hence, at high $SNR$, $d$ represents the number of interference-free signaling dimensions in the network. For the $K$-user IC, with random, time varying channel coefficients drawn from a continuous distribution, the available DoF are found to be $K/2$ [2]. The DoF-optimal achievable scheme is shown to be the interference alignment (IA) scheme [2]. This means that via IA, each user in the $K$-user IC can see half of the available degrees of freedom without interference.

The question that arises is whether each user can see more than half of the space or not? Cooperation between users is perceived to be the most appropriate solution. The $K$-user IC with full cooperation, i.e. if all users cooperate together, turns to be equivalent to a point-to-point MIMO channel with $K$ antennas at each node. The known DoF of this channel is $K$. In practical systems, a certain node can most probably cooperate with a subset of the other nodes rather than all of them. Hence, investigating the available DoF of the $K$-user IC with partial cooperation is of great importance. An outer bound for the DoF of this channel is found in [3], but the exact characterization of the achievable DoF is still an open problem and this is the objective of this paper.

### A. Prior work

In [3], the problem of the $K$-user IC with partial cooperation is studied from a DoF perspective. The term *cooperation order* is defined as the number of users sharing a given message. For the $K$-user IC in which each user has $M - 1$ other cooperating users, i.e. with a cooperation order of $M$, the outer bound of the total available DoF is found to be $KM/(M+1)$. This outer bound is shown to be tight in only two special cases: when $K = 4, M = 2$, and when $K = M + 1$.

The effect of the connectivity of the previous network on the outer bound of DoF is investigated in [4]. Surprisingly, it is shown that missing a link between an interfering transmitter and one of the cooperating receivers, without missing it for the other receivers, will affect the DoF negatively.

### B. Contribution

In this paper, we propose a signal space-based IA scheme that proves the achievability of a total of $KM/(M+1)$ DoF for the $K$-user IC with a cooperation order of $M$ when $K = M + 2$. The scheme proposed consists of a design for the transmit precoding matrices, and an algorithm, SIA algorithm, to be used by receivers in processing the received signals. Due to reciprocity of IA [5], sweeping the roles of transmitters and receivers will lead to the same results.

### C. Significance

Beside the theoretical significance of studying the achievable DoF gained due to partial cooperation, this contribution has a great importance for modern cellular systems, e.g., Long-Term Evolution (LTE), in which base stations are allowed to cooperate together in order to coordinate the interference and increase the capacity of the system, e.g., a coordinated multipoint transmission/reception (CoMP) system [6].

## II. ACHIEVABLE DEGREES OF FREEDOM

*Theorem 1:* For the $K$-user IC with random, time varying channel coefficients drawn from a continuous distribution, and with a cooperation order of $M$, a total of $KM/(M+1)$ degrees of freedom can be achieved if $K = M + 2$.

*Proof:* In the following subsections, a proposed scheme is presented and proved to achieve a total of $KM/(M+1)$ degrees of freedom. The proposed scheme consists of a design for the transmit precoding matrices and a receive processing algorithm which we call Successive Interference Alignment (SIA) algorithm. The following subsections are organized as follows: In subsection A, the system model is illustrated indicating the various elements of the system. These elements include the receive cooperation subsets and the symbol extension used by the proposed scheme. In subsection B, the receive processing algorithm, SIA, is presented which leads to the criteria upon which the transmit precoding matrices are designed. In subsection C, the design for the transmit precoding matrices is shown. Subsection D proves that the proposed scheme achieves $M/(M+1)$ DoF per user, hence, $KM/(M+1)$ DoF for the network when $K = M + 2$.

### A. System Model

Consider the $K$-user IC depicted in Figure 1, in which each transmitter $k$, $k \in \{1, 2, \ldots, K\}$ has independent message $W_k$, and wishes to send it to the respective destination $k$.

An assumption of fully-connected network is used, i.e. the links between all transmitters and receivers are connected. In other words, each receiver has the desired signal from its transmitter in addition to interference from all other $M+1$ transmitters.

The channel coefficients are assumed to be random, time varying and drawn from a continuous distribution. The channel state information (CSI) is assumed to be known globally at all nodes.

Each receiver cooperates with $M - 1$ other receivers. Without loss of generality, the receive cooperation set for receiver $i$ is assumed to be:

$$R_i = \{i, i+1, \ldots, K + (M-1)\}$$

Each node has a single antenna, and uses symbol extension of length $\lambda_n$ [2], where:

$$\lambda_n = \mu_n + M\mu_{n+1}$$

where $\mu_n = n^l, l = K(2M-1), n = 1, 2, 3, \ldots$

We will show that $M\mu_n$ independent, interference-free streams can be achieved per user. Hence, the total number of achievable DoF of the system will be:

$$K\frac{M\mu_n}{\lambda_n} = K\frac{Mn^l}{n^l + M(n+1)^l} \xrightarrow{n \to \infty} K\frac{M}{M+1}$$

Each transmitter $i$ will send its signal in $M\mu_n$ dimensions out of the $\lambda_n$ which is denoted as:

$$X_i = F_n^1 S_i^1 + F_n^2 S_i^2 + \cdots + F_n^M S_i^M$$

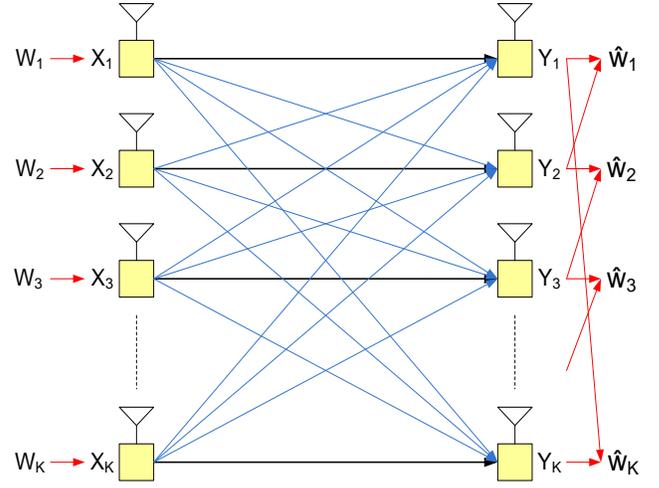

Figure 1. The K-user IC with a cooperation order M=2

where, $X_i$ is the $\lambda_n \times 1$ vector representing the transmitted signal of user $i$, $F_n^j$ is a $\lambda_n \times \mu_n$ matrix containing the transmit directions of the $j^{th}$ part of the signal, and $S_i$ is the $M\mu_n \times 1$ information bearing vector of user $i$.

$$S_i = [S_i^{1T}\ S_i^{2T}\ \ldots\ S_i^{M^T}]^T$$

We use $H_{j,i}$ to denote the $\lambda_n \times \lambda_n$-dimensional diagonal matrices, representing the $\lambda_n$ symbol extension of the channel coefficients between transmitter $i$ and receiver $j$. $Y_j, Z_j$ are used to denote the column vectors of length $\lambda_n$ representing the $\lambda_n$ symbol extension for the received signal and noise terms respectively at receiver $j$.

### B. Successive Interference Alignemtn (SIA) algorithm

The decoder of message $W_k$ has the knowledge of the received signals $Y_k, Y_{k+1}, \ldots, Y_{k+(M-1)}$. Hence, the signals received by this decoder can be represented by $R_k$ shown in equation (1), at the bottom of the page. Note that the noise term is neglected from the equation of $R_k$.

Each row of this equation represents the received signal at one of the $M$ receivers, expressed in $\lambda_n$-dimensional subspace due to the $\lambda_n$ symbol extension. Hence, on a consequence of cooperation, a space of $M\lambda_n$ dimensions is seen by the decoder of message $W_k$. If we can free $M\mu_n$ out of the $M\lambda_n$ dimensions of interference, user $k$ will achieve the required DoF $\frac{M}{M+1}$.

As the network is fully connected, each receiver has the desired signal, the signal of user $k$, in addition to $(K-1)$ interference.

In our proposed scheme, we free $M\mu_n$ dimensions of interference as follows: Firstly, we consider $M$ out of the $(M+1)$ interferers, and process the received signal $R_k$, using SIA algorithm, in order to get $M$ $\lambda_n$-dimensional subspaces, representing the subspaces of the M receivers, with only one

$$R_k = \begin{bmatrix} Y_k \\ Y_{k+1} \\ \vdots \\ Y_{k+M-1} \end{bmatrix} = \begin{bmatrix} H_{k,k} \\ H_{k+1,k} \\ \vdots \\ H_{k+M-1,k} \end{bmatrix} X_k + \begin{bmatrix} H_{k,1} \\ H_{k+1,1} \\ \vdots \\ H_{k+M-1,1} \end{bmatrix} X_1 + \cdots + \begin{bmatrix} H_{k,k-1} \\ H_{k+1,k-1} \\ \vdots \\ H_{k+M-1,k-1} \end{bmatrix} X_{k-1} + \begin{bmatrix} H_{k,k+1} \\ H_{k+1,k+1} \\ \vdots \\ H_{k+M-1,k+1} \end{bmatrix} X_{k+1} + \cdots + \begin{bmatrix} H_{k,K} \\ H_{k+1,K} \\ \vdots \\ H_{k+M-1,K} \end{bmatrix} X_K \quad (1)$$

interferer at each subspace. Hence, at each of these M-subspaces, the interference will occupy $M\mu_n$-dimensional subspace. This interference subspace can be defined by $C_k[F_n^1\ F_n^2 \dots F_n^M]$, where $C_k$ is a general coefficient; as will be illustrated in the SIA algorithm. The remaining interferer and at each subspace are then aligned to the subspace $C_k[F_{n+1}^1\ F_{n+1}^2 \dots F_{n+1}^M]$ which is a $M\mu_{n+1}$-dimensional subspace determined by the previous M interferers after processing. At each subspace, $(M-1)\mu_n$ dimensions of the desired signal are also aligned into the interference subspace. Hence, for the M subspaces, the interference will occupy $M\mu_{n+1}$ dimensions out of the $M.\lambda_n$ dimensions, leaving $M\mu_n$ dimensions free of interference for the desired signal. Therefore, the proposed scheme consists of two parts: The SIA algorithm which is used to process the received signal $R_k$, and the design of the transmit precoding matrices, $F_n^j$.

Before presenting the SIA algorithm for the general case of the $K$-user IC with a cooperation order of $M = K - 2$, we start with the case of the 5-user IC with a cooperation order of $M=3$. Considering the previous system model, the decoder of message $W_k$ has the knowledge of $Y_k, Y_{k+1}, Y_{k+2}$. Hence, they can be represented as:

$$R_k = \begin{bmatrix} Y_k \\ Y_{k+1} \\ Y_{k+2} \end{bmatrix} = \begin{bmatrix} H_{k,k} \\ H_{k+1,k} \\ H_{k+2,k} \end{bmatrix} X_k + \begin{bmatrix} H_{k,k-1} \\ H_{k+1,k-1} \\ H_{k+2,k-1} \end{bmatrix} X_{k-1}$$
$$+ \begin{bmatrix} H_{k,k+1} \\ H_{k+1,k+1} \\ H_{k+2,k+1} \end{bmatrix} X_{k+1} + \begin{bmatrix} H_{k,k+2} \\ H_{k+1,k+2} \\ H_{k+2,k+2} \end{bmatrix} X_{k+2}$$
$$+ \begin{bmatrix} H_{k,k+3} \\ H_{k+1,k+3} \\ H_{k+2,k+3} \end{bmatrix} X_{k+3} \quad (2)$$

Considering the interferers $X_{k-1}, X_{k+1}, X_{k+2}$, we want to process $R_k$ in order to have 3 subspaces with only one of these 3 interferers at each subspace. This is done in 2 steps.

In the first step, the component of the interferer $X_{k-1}$ from the first subspace, the first row in the equation of $R_k$, is used to cancel the component of the interferer $X_{k-1}$ from the second subspace. Similarly, the components of the interferer $X_{k+1}, X_{k+2}$ from the second and third subspaces are used to cancel the components of the interferer $X_{k+1}, X_{k+2}$ from the third and first subspaces respectively. The following processing represents this step:

$$R_k^{[1]} = \begin{bmatrix} -H_{k+2,k+2} & 0 & H_{k,k+2} \\ H_{k+1,k-1} & -H_{k,k-1} & 0 \\ 0 & H_{k+2,k+1} & -H_{k+1,k+1} \end{bmatrix} R_k \quad (3)$$

The subspaces are now defined by $R_k^{[1]}$ where:

$$R_k^{[1]} = \begin{bmatrix} G_k^{[1]} \\ G_{k+1}^{[1]} \\ G_{k+2}^{[1]} \end{bmatrix} X_k + \begin{bmatrix} T_{k,k}^{[1]} \\ 0 \\ T_{k+2,k}^{[1]} \end{bmatrix} X_{k-1} + \begin{bmatrix} T_{k,k+1}^{[1]} \\ T_{k+1,k+1}^{[1]} \\ 0 \end{bmatrix} X_{k+1}$$
$$+ \begin{bmatrix} 0 \\ T_{k+1,k+2}^{[1]} \\ T_{k+2,k+2}^{[1]} \end{bmatrix} X_{k+2} + \begin{bmatrix} T_{k,k+3}^{[1]} \\ T_{k+1,k+3}^{[1]} \\ T_{k+2,k+3}^{[1]} \end{bmatrix} X_{k+3} \quad (4)$$

where

$$G_k^{[1]} = H_{k,k+2}H_{k+2,k} - H_{k,k}H_{k+2,k+2}$$
$$G_{k+1}^{[1]} = H_{k+1,k-1}H_{k,k} - H_{k,k-1}H_{k+1,k}$$
$$G_{k+2}^{[1]} = H_{k+2,k+1}H_{k+1,k} - H_{k+1,k+1}H_{k+2,k}$$

Similarly, the remaining coefficients can be derived. Note that the superscript [1] indicates that these are the coefficients after the first step.

In the second step, the component of the interferer $X_{k-1}$ from the first subspace, the first row in the equation of $R_k^{[1]}$, is used to cancel the component of the interferer $X_{k-1}$ from the third subspace. However, this processing induces a component for the interferer $X_{k+1}$ at the third subspace. Hence, the component of the interferer $X_{k+1}$ from the second subspace will be used again to cancel the induced component at the third subspace. Similarly, the components of the interferer $X_{k+1}, X_{k+2}$ from the first and second subspaces respectively are cancelled. The following processing represents this step:

$$R_k^{[2]} =$$
$$\begin{bmatrix} -H_{k+1,k+1}H_{k+2,k+2} & H_{k,k+1}H_{k+2,k+2} & -H_{k,k+1}H_{k+1,k+2} \\ -H_{k+2,k-1}H_{k+1,k+2} & -H_{k+2,k+2}H_{k,k-1} & H_{k+1,k+2}H_{k,k-1} \\ H_{k+2,k-1}H_{k+1,k+1} & -H_{k+2,k-1}H_{k,k+1} & -H_{k,k-1}H_{k+1,k+1} \end{bmatrix} R_k^{[1]} \quad (5)$$

The subspaces are now defined by $R_k^{[2]}$ where:

$$R_k^{[2]} = \begin{bmatrix} G_k^{[2]} \\ G_{k+1}^{[2]} \\ G_{k+2}^{[2]} \end{bmatrix} X_k + \begin{bmatrix} T_{k,k}^{[2]} \\ 0 \\ 0 \end{bmatrix} X_{k-1} + \begin{bmatrix} 0 \\ T_{k+1,k+1}^{[2]} \\ 0 \end{bmatrix} X_{k+1}$$
$$+ \begin{bmatrix} 0 \\ 0 \\ T_{k+2,k+2}^{[2]} \end{bmatrix} X_{k+2} + \begin{bmatrix} T_{k,k+3}^{[2]} \\ T_{k+1,k+3}^{[2]} \\ T_{k+2,k+3}^{[2]} \end{bmatrix} X_{k+3} \quad (6)$$

In $R_k^{[2]}$, we will find that $T_{k,k}^{[2]} = T_{k+1,k+1}^{[2]} = T_{k+2,k+2}^{[2]}$.

For the general case, $M - 1$ steps are needed. In each step, one interferer is cancelled from one of the M subspaces. The SIA algorithm has a recursive nature as shown from second step. The following algorithm is the SIA algorithm for the general $K$-user IC with a cooperation order of $M$.

**Algorithm 1** Successive Interference Alignment

1: **Inputs**: $R = [T_{m,n}], \forall m \in \{1,2,..,M\}, n \in \{1,2,..,K\}$
for $D = 1$ to $M - 1$ // M interference will be cancelled in each step.
2:     for $k = 1$ to $M$ //iterate for all the M subspaces.
3:         $R = \text{Align}(R, k, k + D)$
4:     **end for**
5: **end for**

1: **Function**: Align($R, i, j$)

   // It uses the $i^{th}$ interference of the $i^{th}$ subspace in order to cancel the $i^{th}$ interference of the $j^{th}$ subspace.

2: Subspace $j = T_{i,i}$. Subspace $j - T_{j,i}$. Subspace $i$.

   // where subspace $i$ and $j$ are the $i^{th}$ and $j^{th}$ rows of matrix $R$ respectively.

3: If ($j - i > 1$), then R= Align($R, i + 1, j$).

4: **End function**

After using the SIA algorithm in processing the received signals $R_k$, they can be expressed as in equation (7), in the bottom of the page, for the general case of $K = (M + 2)$ users and a cooperation order of M. $G_{k,i}^{[M-1]}$ represents the coefficient of the desired signal at the $i^{th}$ subspace, after $M - 1$ processing iterations of the SIA algorithm at the decoder of message k. $T_k^{[M-1]}$ is the coefficient of the 1st $M$ interference signals at the $M$ subspaces, after $M - 1$ iterations at the decoder of message $k$. $T_{k,i}^{[M-1]}$ represents the coefficient of the remaining interference at the $i^{th}$ subspace, after $M - 1$ processing iterations at the decoder of message $k$.

*C. Design of the transmit precoding matrices*

The interference is now defined by the subspace in equation (8) at the bottom of the page.

where, span(A) means the space spanned by the columns of matrix A. Note that all of these coefficients are after using the SIA algorithm, but we remove the superscript $M - 1$ for simplicity.

Now, we want to design the transmit precoding matrices $F_n^{\ j}$, in order to satisfy the condition in equation (9), at the bottom of the page. If this condition is satisfied, the interference will be defined by the subspace:

$$\text{span}\left(\begin{bmatrix} T_k F_{n+1}^{\ 1} & T_k F_{n+1}^{\ M} & 0 & 0 \\ 0 & 0 & 0 & 0 \\ 0 & \cdots & 0 & \ddots & \vdots & \cdots & \vdots \\ \vdots & & \vdots & & 0 & & 0 \\ 0 & & 0 & & T_k F_{n+1}^{\ 1} & & T_k F_{n+1}^{\ M} \end{bmatrix}\right)$$

In order to do that, the following conditions must be satisfied:

$$T_{k,i} F_n^{\ j} \prec T_k F_{n+1}^{\ j}$$
$$G_{k,r} F_n^{\ j} \prec T_k F_{n+1}^{\ j}$$
$$\forall k \in \{1,2,\ldots,K\}, i,j,r \in \{1,2,\ldots,M\}, r \neq j$$

where $A \prec B$ means that the set of column vector of matrix $A$ is a subset of the set of column vectors of matrix $B$.

Hence, for each precoding matrix $F_n^{\ j}$, there are $K(2M - 1)$ such conditions, and we need to choose $\mu_n$ column vectors for $F_n$ and $\mu_{n+1}$ column vectors for $F_{n+1}$. Let $w$ be a $\lambda_n \times 1$ column vector, $w = [1 \ 1 \ldots 1]^T$. $F_n$ and $F_{n+1}$ are chosen to be:

$F_n^{\ j}$
$$= \left\{\left(\prod_{k \in A}\left(\prod_{i \in I}(T_{k,i})^{\alpha_i} \cdot \prod_{r \in R}(G_{k,r})^{\beta_r}\right) \cdot T_k^{P-\sum_{i \in I}(\alpha_i)-\sum_{r \in R}(\beta_r)}\right)w\right\}$$

$\forall \ \alpha_i, \beta_r \in \{0,1,\ldots,n-1\}, P = (2M - 1)n, A = \{1,2,\ldots,K\},$

$I = \{1,2,\ldots,M\}, R = \{1,2,\ldots,j-1,j+1,\ldots,M\}$

Using these transmit precoding matrices, the signal and interference subspaces can be described by the following matrix:

$$\begin{bmatrix} G_{k,1}F_n^{\ 1} & 0 & T_k F_{n+1}^{\ 1} & T_k F_{n+1}^{\ M} & 0 & 0 \\ 0 & 0 & 0 & 0 & 0 & 0 \\ 0 & \ddots & 0 & 0 & \cdots & 0 & \ddots & \vdots & \cdots & \vdots \\ \vdots & & \vdots & \vdots & & \vdots & & 0 & & 0 \\ 0 & G_{k,M}F_n^{\ M} & 0 & 0 & T_k F_{n+1}^{\ 1} & T_k F_{n+1}^{\ M} \end{bmatrix}$$

(10)

$$R_k^{[M-1]} = \begin{bmatrix} G_{k,1}^{[M-1]} \\ G_{k,2}^{[M-1]} \\ G_{k,3}^{[M-1]} \\ \vdots \\ G_{k,M}^{[M-1]} \end{bmatrix} X_k + \begin{bmatrix} T_k^{[M-1]} \\ 0 \\ 0 \\ \vdots \\ 0 \end{bmatrix} X_1 + \begin{bmatrix} 0 \\ T_k^{[M-1]} \\ 0 \\ \vdots \\ 0 \end{bmatrix} X_2 + \cdots + \begin{bmatrix} 0 \\ 0 \\ \vdots \\ 0 \\ T_k^{[M-1]} \end{bmatrix} X_M + \begin{bmatrix} T_{k,1}^{[M-1]} \\ T_{k,2}^{[M-1]} \\ \vdots \\ T_{k,M-1}^{[M-1]} \\ T_{k,M}^{[M-1]} \end{bmatrix} X_{M+1} \quad (7)$$

$\underbrace{\phantom{XXXXXX}}_{\text{The desired signal } X_k} \quad \underbrace{\phantom{X}}_{\text{The (M+1) interference signals(all signals excluding } X_k)}$

$$\text{span}(I) = \text{span}\left(\begin{bmatrix} T_k F_n^{\ 1} & T_k F_n^{\ M} & 0 & 0 & T_{k,1}F_n^{\ 1} & T_{k,1}F_n^{\ M} \\ 0 & 0 & 0 & 0 & T_{k,2}F_n^{\ 1} & T_{k,2}F_n^{\ M} \\ 0 & \cdots & 0 & \ddots & \vdots & \cdots & \vdots & \vdots & \cdots & \vdots \\ \vdots & & \vdots & & 0 & & 0 & T_{k.M-1}F_n^{\ 1} & T_{k.M-1}F_n^{\ M} \\ 0 & & 0 & T_k F_n^{\ 1} & T_k F_n^{\ M} & T_{k,M}F_n^{\ 1} & T_{k,M}F_n^{\ M} \end{bmatrix}\right) \quad (8)$$

$$\text{span}\left(\begin{bmatrix} 0 & G_{k,1}F_n^{\ 1} & G_{k,1}F_n^{\ M} \\ G_{k,2}F_n^{\ 1} & 0 & G_{k,2}F_n^{\ M} \\ G_{k,3}F_n^{\ 1} & G_{k,3}F_n^{\ 1} & \cdots & G_{k,3}F_n^{\ M} \\ \vdots & \vdots & & \vdots \\ G_{k,M}F_n^{\ 1} & G_{k,M}F_n^{\ 1} & & 0 \end{bmatrix}\right) \cup \text{span}(I) \prec \text{span}\left(\begin{bmatrix} T_k F_{n+1}^{\ 1} & T_k F_{n+1}^{\ M} & 0 & 0 \\ 0 & 0 & 0 & 0 \\ 0 & \cdots & 0 & \ddots & \vdots & \cdots & \vdots \\ \vdots & & \vdots & & 0 & & 0 \\ 0 & & 0 & & T_k F_{n+1}^{\ 1} & & T_k F_{n+1}^{\ M} \end{bmatrix}\right) \quad (9)$$

In order to prove that the desired transmitted signal, which is transmitted over $M\mu_n$ dimensions, can be decoded interference-free, it is sufficient to prove that the previous $M\lambda_n \times M\lambda_n$ matrix (10) is full rank.

In order to prove that the matrix (10) is full rank, it is sufficient to prove that the following conditions are satisfied:

$$\text{rank}([G_{k,1}F_n^{\ 1}\ T_kF_{n+1}^{\ 1}\ ...\ T_kF_{n+1}^{\ M}]) = \lambda_n$$
$$\text{rank}([G_{k,2}F_n^{\ 2}\ T_kF_{n+1}^{\ 1}\ ...\ T_kF_{n+1}^{\ M}]) = \lambda_n$$
$$\vdots$$
$$\text{rank}([G_{k,M}F_n^{\ M}\ T_kF_{n+1}^{\ 1}\ ...\ T_kF_{n+1}^{\ M}]) = \lambda_n$$

It can be shown that each column in the previous matrices is the resultant of an addition of a number of terms. Each term is the multiplication of a number of channel coefficients. These terms are different from one column to another. Now, as the channel coefficients are random and drawn from a continuous distribution, the columns of the previous matrices can be shown to be independent almost surely. Hence, the previous conditions are satisfied, and the matrix (10) is full rank for the case $K = M + 2$ almost surely.

### III. CONCLUSION AND ON-GOING WORK

In this paper, we considered the $K$-user IC with partial cooperation. If each user has $M - 1$ other cooperating users, i.e. for a cooperation order of $M$, we showed that a total of $KM/(M + 1)$ DoF can be achieved when $K = M + 2$. We proposed a signal space-based IA scheme that achieves the maximum available DoF. It consists of a design for the transmit precoding matrices, and a processing algorithm, SIA, in order to be used by each receiver in processing the received signals. In our current work, we attempt studying the achievable degrees of freedom problem for the channel mentioned when $K$ is greater than $M + 2$.